 \documentclass[12pt,preprint]{aastex}

\newcommand{\etal}{\it{et al.}\rm}

\shorttitle{Cool Hypergiants}
\shortauthors{Schuster \etal}

\begin{document}

\title{The Circumstellar Environments of NML Cyg and the Cool Hypergiants}

\author{Michael T. Schuster\altaffilmark{1} and Roberta M. Humphreys} 
\affil{School of Physics and Astronomy, University of Minnesota, Minneapolis, 
MN 55455}
\email{mschuster@cfa.harvard.edu and roberta@aps.umn.edu}

\author{and}

\author{Massimo Marengo} 
\affil{Harvard-Smithsonian Center for Astrophysics, 60 Garden St., Cambridge, 
MA 02138}
\email{mmarengo@cfa.harvard.edu}

\altaffiltext{1}{Smithsonian Astrophysical Observatory Predoctoral Fellow at 
Harvard-Smithsonian Center for Astrophysics, 60 Garden St., Cambridge, MA 
02138}

\begin{abstract}

We present high-resolution {\it HST} WFPC2 images of compact nebulosity 
surrounding the cool M--type hypergiants NML~Cyg, VX~Sgr and S~Per. The 
powerful OH/IR source NML~Cyg exhibits a peculiar bean-shaped asymmetric 
nebula that is coincident with the distribution of its H$_{2}$O vapor masers. 
We show that NML~Cyg's circumstellar envelope is likely shaped by 
photo-dissociation from the powerful, nearby association Cyg OB2 inside the 
Cygnus X superbubble. The OH/IR sources VX~Sgr and S~Per have marginally 
resolved envelopes. S Per's circumstellar nebula appears elongated in a NE/SW 
orientation similar to that for its OH and H$_{2}$O masers, while VX Sgr is 
embedded in a spheroidal envelope. We find no evidence for circumstellar 
nebulosity around the intermediate--type hypergiants $\rho$~Cas, HR~8752, 
HR~5171a, nor the normal M--type supergiant $\mu$~Cep. We conclude that there 
is no evidence for high mass loss events prior to 500-1000~yrs ago for these 
four stars.

\end{abstract}


\section{Introduction}

A few highly unstable, very massive stars lie on or near the empirical upper 
luminosity boundary in the HR diagram 
\citep[and Figure 1, this paper]{humphreys79, humphreys94, humphreys83}. These 
include the Luminous Blue Variables, the cool hypergiants, and even rarer 
objects, all related by high mass loss phenomena, sometimes violent, which may 
be responsible for the existence of the upper boundary. In this paper, we use 
the term `cool hypergiant' for the stars that lie just below this upper 
envelope with spectral types ranging from late A to M. The cool hypergiants 
represent a very short-lived evolutionary stage, with time scales of only a 
few~$\times~10^{5}$ years, or less, as a red supergiant (RSG) and possibly as 
short as a few thousand years in transit from the main sequence to the red 
supergiant stage and back again to warmer temperatures. Very high mass loss 
rates have been measured for many of these stars. Recent observations of 
two of these stars, the warm OH/IR post-RSG IRC~+10420 and the peculiar OH/IR 
M--type supergiant VY~CMa, have yielded surprising results about their 
circumstellar environments, including evidence for asymmetric ejections and 
multiple high mass loss events 
\citep{humphreys97,smith01,humphreys02,humphreys05}.

\citet{dejager98} has suggested that most if not all of the intermediate 
temperature hypergiants are post-RSGs. In their post-RSG blueward evolution 
these very massive stars enter a temperature range (6000--9000~K) with 
increased dynamical instability, a semi-forbidden region in the HR diagram, 
that he called the {\it ``yellow void''}, where high mass loss episodes occur. 
Based on information from our {\it HST}/STIS spectra of IRC~+10420, we 
\citep{humphreys02} demonstrated that its wind is optically thick, and 
therefore concluded that its observed spectral changes are not due to rapid 
evolution, although the star may be about to shed its dense wind, cross the 
yellow void, and emerge as a warmer star. However, in contrast with 
IRC~+10420, the evolutionary state of most of the cool hypergiants is not 
known. They may be evolving toward the RSG region or back to the blue side of 
the HR diagram after having lost considerable mass as RSGs. The 
post-RSG state for some of these stars (i.e $\rho$~Cas) is supported by a 
substantial overabundance of N and Na \citep{takeda94,eleid95}.

To better understand the evolution of cool, evolved stars near the upper 
luminosity boundary and the mass loss mechanisms that dominate the upper HR 
diagram, we obtained high resolution multi-wavelength images with {\it 
HST}/WFPC2 of seven of the most luminous known evolved stars - the M--type 
hypergiants, $\mu$~Cep (M2e~Ia), S~Per (M3-4e~Ia), NML~Cyg (M6~I), and VX~Sgr 
(M4e~Ia--M9.5~I), and the intermediate--type (F and G--type) hypergiants, 
$\rho$~Cas (F8p~Ia), HR~8752 (G0-5~Ia) and HR~5171a (G8~Ia). The presence or 
lack of fossil shells, bipolar or equatorial ejecta, and other structures in 
their circumstellar environments will be a record of their current and prior 
mass loss episodes and provide clues to their evolutionary history. These 
stars were selected on the basis of their infrared emission, strong molecular 
emission, or peculiar spectroscopic variations to give us a snapshot of 
different steps in their evolution across the top of the HR Diagram. In the 
next section we describe the observations and data reduction procedures. In 
sections 3 and 4 we present the resulting images for each of these stars and 
their circumstellar environments. In the last section we discuss the 
implications for these stars' mass loss histories and evolutionary states.

\section{Observations and Analysis}

The multi-wavelength images of these seven very luminous cool stars were 
obtained in late 1999 and early 2000 with the WFPC2 Planetary Camera on {\it 
HST}. The observations were planned to search for material close to the star 
as well as more distant nebulosity. Since we are interested in imaging faint 
ejecta associated with some relatively bright stars, we used a range of 
exposure times in each filter with the shortest to avoid saturation and 
minimize bleeding and the longest to detect faint emission. A variety of 
filters were chosen to look for changes, if any, in the circumstellar material 
at different wavelengths. For the OH/IR M--type hypergiants we used broad band 
Johnson-Cousins and medium band Str\"{o}mgren filters (see Table~\ref{data}). 
We used a combination of narrow band forbidden line filters for the 
intermediate--type hypergiants, as well as the normal M--type supergiant 
$\mu$~Cep. These narrow band filters were chosen mainly as continuum filters 
to limit the collected flux for the extremely visually bright stars. All but 
S~Per were observed in H$_{\alpha}$. We dithered our exposures with 2.5~pixel 
shifts in each direction on the detector in order to increase the standard PC 
sampling of 0\farcs04555~pix$^{-1}$ by $2\times2$ to 0\farcs02277~pix$^{-1}$. 
The observations are summarized in Table \ref{data}.

Prior to co-addition, the images were processed with STScI's standard 
calibration using the most recent reference files. Multiple, dithered 
exposures allowed us to remove cosmic rays, bad pixels, and other effects 
during co-addition. We combined our images with the IRAF/STSDAS software 
package DITHER which uses drizzling (to recover image resolution from the 
pixel response of the camera while preserving photometric accuracy, 
\citealt{koekemoer00,fruchter02}) and cross-correlation of sources in the 
field for relative alignment between dithered images. An added benefit of the 
DITHER package is that a short, underexposed image can be scaled by the 
relative exposure times and patched onto an overexposed image during the 
co-addition stage. We generally followed the process described by 
\citet{humphreys97} for patching underexposed point-spread functions (PSF) 
onto overexposed WFPC2 images. To determine the area in the long exposures to 
patch the PSFs, we masked the pixels that were saturated, showed signs of 
bleeding, or had a large percent difference from a median image. By doing this 
we were able to directly patch a non-saturated PSF onto the co-added image 
during the drizzling process. Elsewhere on a given image the short exposure is 
weighted significantly less than the long exposures so that its pixels are 
essentially ignored. The resulting ``PSF-patched'' image has both a higher 
resolution and a larger dynamic range increasing the likelihood of detecting 
bright material near the star and faint material far from the star. Our 
patched images have dynamic ranges for the PSF of $\sim1.4\times10^{3}$ up to 
$3.3\times10^{4}$, compared to $\sim2300$, which is the maximum for a single 
non-saturated WFPC2 PC image where the background is dominated by read 
noise\footnote{This compares a PSF peak near saturation at 3500DN to a 3 sigma 
feature where the read noise of the CCD is 7e$^{-}$ and the gain is 
14e$^{-}$/DN.}.

To suppress the prominent {\it HST} diffraction spikes and rings of the bright 
stars in our program we used TinyTim\footnote{www.stsci.edu/software/tinytim} 
PSFs for both subtraction and deconvolution (see \citealt{biretta00}, ch.~7). 
We subsampled the PSFs by $2\times2$ to match the sampling of our drizzled 
images. We compared our images to several versions of these PSFs by including 
different combinations of the following: filter throughput, chip position, 
telescope focus, charge diffusion and the rms jitter of the spacecraft. For 
our medium and broad band filter observations we also included a model 
spectrum and an estimate of the interstellar reddening which is significant 
for many of our targets. The brightness of these sources lead to very high 
signal-to-noise PSFs in almost all cases. Consequently, differences between 
our observations and the TinyTim PSFs are apparent in both the subtraction and 
deconvolution residuals. These residual artifacts generally appear as rings or 
remnants of the bright diffraction spikes. We judged the quality of our 
subtractions and deconvolutions based on their chi-square fit (deconvolutions 
only), comparison with other point sources in the image, and the relative 
residuals between sources. Our ability to detect faint material near the star 
is limited by the quality of the PSF, and so we were careful to preserve PSF 
fidelity. We used both interpolation and super-sampled PSFs (5$\times$ higher 
sampling) separately to mitigate the effects of pixel phase in the shape of 
the PSF. Together these steps helped to minimize the subtraction and 
deconvolution residuals. NML~Cyg, VX~Sgr and S~Per are extended {\it and} do 
not exhibit point-sources, and so PSF subtraction was not very illustrative. 
We deconvolved the PSF from our images with the IRAF/STSDAS task LUCY which 
uses the Richardson-Lucy technique \citep{richardson72,lucy74,snyder90}. Our 
images of NML~Cyg, VX~Sgr and S~Per appeared sharper after deconvolution.

The final co-added, cleaned and PSF-patched images for the OH/IR M supergiants 
NML~Cyg, VX~Sgr and S~Per (Figures \ref{nml1}, \ref{vxsgr} and \ref{sper}) 
show the presence of extended nebulosity, and are discussed individually in 
the next section. However, we find no evidence for extended structure around 
the intermediate--type hypergiants $\rho$~Cas, HR~8752, and HR~5171a and the 
normal M supergiant $\mu$~Cep. We will discuss the implications of these 
detections and non-detections on the evolutionary status of these stars in 
Sections \ref{intermediate} and \ref{massloss}.

\section{The Circumstellar Environments of the M--type Hypergiants}{\label{mtype}}

NML~Cyg, VX~Sgr and S~Per are powerful OH/IR supergiants with strong maser 
emission from OH, H$_{2}$O, and SiO plus large infrared excess radiation. Our 
images, the highest resolution images of these stars in visible light to date, 
show circumstellar material surrounding all three, but NML~Cyg is the most 
intriguing with its asymmetric bean-like shape and the probable interaction 
between its strong wind and its interstellar environment due to its proximity 
to the Cyg~OB2 association. We find no evidence in our images for 
circumstellar material around the normal M supergiant $\mu$~Cep (see 
Table~\ref{limitsI}), although like other RSGs $\mu$~Cep has a 
10{\micron} silicate emission feature due to dust. Weak K~I emission has also 
been detected in its wind \citep{mauron97} and in the winds of several other 
red giants and supergiants \citep{guilain96}.

\subsection{NML~Cyg}

The powerful OH/IR source NML~Cyg (M6~I) is approximately 1.7~kpc from the sun 
and $\sim100$~pc from the large association Cyg~OB2 in the X-ray emitting 
Cygnus~X superbubble \citep{humphreys78,morris83,knodlseder03}. This distance 
places it near the empirical upper luminosity boundary for RSGs 
with a luminosity of $5\times10^5~L_{\sun}$ ($M_{bol}\sim-9.5$) and a mass 
loss rate of $6.4\times10^{-5}~M_{\sun}yr^{-1}$ \citep{hyland72,morris83}. Our 
WFPC2 images show that NML~Cyg has a very obvious circumstellar nebula with a 
peculiar asymmetric shape; Figures \ref{nml1} and \ref{nml2} (deconvolved) 
show our F555W image. In addition to the asymmetric component to the nebula, 
there is a bright area that is roughly spherical and is most likely the 
location of the embedded star. There is little difference among the F555W, 
F656N and F675W images, but NML~Cyg is too faint at blue wavelengths to be 
detected in our F439W exposure. There is no noticeable shift in position, nor 
any change in appearance of the obscured star with wavelength. The nebula also 
appears to be slightly lopsided about a line of symmetry that runs WNW/ESE, 
though we are unable to differentiate between instrumental effects and any 
true asymmetry at that level of signal. The nebulosity appears more diffuse to 
the North, though it is possible that this is due to blurring by a diffraction 
spike in the PSF. This blurring is still evident in the deconvolved image, and 
is not unexpected since the TinyTim PSF does not model the diffraction spikes 
well. The two columns of brighter pixels just to the South of the embedded 
star (bright spot) may be a result of a small amount of bleeding in the CCD 
(see Figure~\ref{nml1}).

\citet{monnier97} and \citet{blocker01} find that the dust around NML~Cyg may 
be concentrated in one or more shells based on their infrared interferometry 
observations, with an inner radius of about 100~mas for the main dust shell. 
Both authors conclude that multiple shells are necessary to fit the observed 
profiles. Their outermost shells coincide with the outer edge of the visible 
nebula seen in our images (at a radius of $\sim0\farcs3$). Thus, it is likely 
that the emission that they detected is from the dust responsible for 
scattering the stellar light seen in our images.

Observations of NML~Cyg by \citet{masheder74} show that the 1612 MHz OH masers 
extend up to $2\arcsec$ from the star, and that the emission is elongated along 
a NW/SE axis with a position angle of about $150\degr$ which has been 
confirmed by \citet{benson79}, \citet{diamond84}, and others. The H$_{2}$O 
masers are much closer to the embedded star and show an asymmetric 
distribution to the SE \citep{richards96} similar in size to the optical 
reflection nebula (see Figure~\ref{nml2}). \citet{richards96} have suggested 
that the NW/SE spatial distribution of the H$_{2}$O vapor masers may indicate 
a bipolar outflow with p.a. $132\degr$. Recent SiO ground state observations 
indicate a similar NW/SE axis, and an asymmetry in the emission towards the SE 
\citep[and private communication]{boboltz04}. It is also possible that the 
maser emission is tracing an asymmetric, episodic outflow that may be 
reminiscent of the arcs and other structures seen in the circumstellar nebula 
surrounding VY~CMa \citep{smith01}. Given their asymmetric, one-sided 
distribution within NML~Cyg's nebula, one possible explanation is that the 
masers are protected by the star's envelope from Cyg~OB2's radiation 
as discussed below.

There are remarkable similarities between the asymmetric envelope that we see 
and the much more distant ($\sim30\arcsec$ from the star) 21~cm ionized 
hydrogen (H~II) contours around NML~Cyg observed by \citet{habing82}. 
\citet{morris83} showed that the asymmetric ``inverse'' H~II region was the 
result of the interaction of a spherically symmetric, expanding wind from 
NML~Cyg and photo-ionization from plane parallel Lyman continuum photons from 
the luminous, hot stars in the nearby association Cyg~OB2 (see Figures 1 and 2 
in \citeauthor{morris83}). They described the interaction by balancing the 
incident ionizing flux against recombinations and atoms lost from the star. 
The 21~cm ``emission measure'' along the computed ionization surface 
(projected along the observer's line-of-sight) convolved with the telescope 
beam from \citeauthor{habing82} produced normalized contours that closely 
matched the observed H~II region (see their Figure~2). \citeauthor{morris83} 
demonstrated that the strength of the Lyman continuum flux from Cyg~OB2 and 
the density of atomic hydrogen around NML~Cyg are sufficient to produce the 
observed 21~cm emission.

The presence of ionized hydrogen surrounding an M supergiant like NML~Cyg was 
somewhat of an enigma. To explain its presence, \citeauthor{morris83} 
suggested that the molecular material in the wind is photo-dissociated closer 
to the star so that it does not shield the atomic hydrogen from the ionizing 
photons (from Cyg~OB2) farther out. They estimated the number of dissociating 
photons emitted by Cyg~OB2 as roughly equal to or greater than the number of 
Lyman continuum photons ($I_{L_{\alpha}}\sim7\times10^{8}$~cm$^{-2}$s$^{-1}$, 
\citealt{morris83}, and recently increased by a factor of 2 by 
\citealt{knodlseder03}). The dissociation boundaries for H$_{2}$O, OH, and 
other molecules are expected to be closer to NML~Cyg than the ionization 
surface. In other words, within the Cygnus~X superbubble, UV photons are able 
to survive the roughly 100~pc journey from Cyg~OB2 to NML~Cyg. The Lyman 
continuum photons, $\geq13.6$~eV, will be attenuated from the UV flux by the 
ionization front farther out from NML~Cyg leaving the less energetic photons 
to dissociate the molecular material closer in.

Our images show circumstellar material much closer to NML~Cyg than the 
surrounding H~II region and coincident with the water masers (see 
Figure~\ref{nml2}), as well as SiO masers, suggesting that we are likely 
imaging the photo-dissociation boundaries. We propose that the shape of the 
envelope seen in our WFPC2 images is the result of the interaction between the 
molecular outflow from NML~Cyg and the near--UV continuum flux from Cyg~OB2, 
i.e. analogous to an ``inverse Photo-Dissociation Region'' (PDR). To test our 
hypothesis we assume that the expansion of the envelope is spherically 
symmetric and that the near--UV flux, $I_{NUV}$ (cm$^{-2}$sec$^{-1}$), from 
Cyg~OB2 is plane-parallel. We assume that the fraction of photons that 
photo-dissociate is $f$, with the net dissociating flux equal to $I_{NUV}f$. 
We then calculate the relationship between the radial distance to the 
dissociation surface $r$ and position angle $\beta$ from the direction of the 
incident flux by balancing the incoming radiation against the molecular 
outflow:

\begin{eqnarray}
I_{NUV}f\sin{(\beta+\phi)} = \dot{N}\sin{\phi}/4{\pi}r^{2}
\end{eqnarray}

\noindent where $\dot{N}$ (s$^{-1}$) is the number of molecules per second 
lost from the star, and $\phi$ is the angle between the radius vector and the 
tangent to the dissociation surface ($\leq\pi/2$) (see Figure \ref{surface}). 
The $\sin{(\beta+\phi)}$ factor projects the incident flux onto the 
dissociation surface. Rearranging eq. (1), and noting that 
$\cot{\phi}=d\ln{r}/d\beta$ we obtain:

\begin{eqnarray}
I_{NUV}f4{\pi}r^{2}/\dot{N} & = & [\cos{\beta} + (d\ln{r}/d\beta)\sin{\beta}]^{-1}
\end{eqnarray}

\noindent with the initial condition $d\ln{r}/d\beta=0$ at $\beta=0$. We 
insert in eq. (2) the minimum separation between the star and the dissociation 
surface $r_{o}=\sqrt{\dot{N}/4{\pi}I_{NUV}f}$ and find the solution:

\begin{eqnarray}
r & = & \frac{r_{o}}{\cos{(\beta/2)}}
\end{eqnarray}

To estimate the rate that molecules are lost from NML~Cyg we use 
$\dot{N}=4{\pi}r^{2}_{o}I_{NUV}f$. We take the dissociating near--UV flux 
below 13.6~eV, $I_{NUV}f$ with $f\sim1/2$, to be of the order of 
$\sim~10^{9}$~cm$^{-2}$s$^{-1}$ based on \citet{morris83}'s reasoning and 
estimated Lyman continuum flux \citep[revised by][]{knodlseder03}. In order to 
get an estimate for the size scale for the dissociation surfaces, $r_{o}$, we 
use the radial profile shown in Figure~\ref{nmlfit}. The profile is taken as a 
$60\degr$ wedge in the direction of Cyg~OB2. We find that two components are 
necessary to fit the observed shape of the profile, one for the embedded star 
(a gaussian, FWHM=4.7~pix, convolved with the PSF) and another for the 
asymmetric nebula (a gaussian, FWHM=9.4~pix, convolved with the PSF). We adopt 
$r_{o}\sim0\farcs1$, the FWHM/2 of the fit to the asymmetric nebula, as the 
size scale for the dissociation boundaries. This corresponds to about 170~AU 
at NML~Cyg's distance. From this we find a good order-of-magnitude estimate of 
$\dot{N}\sim~10^{41}~s^{-1}$, or about $5\times10^{-8}~M_{\sun}yr^{-1}$ 
assuming an average mass of $\sim20m_{H}$. This may be a lower limit since we 
have neglected the increased density of the maser emission.

The solution for the dissociation surface in eq. (3) is plotted in 
Figure~\ref{surface} as it would be seen edge-on (solid curve), along with a 
projection of the surface that is inclined from our line-of-sight by $60\degr$ 
(dot-dashed curve). If NML~Cyg and Cyg~OB2 are interacting, the inclination of 
the envelope will be roughly in the range $60 - 120\degr$ (edge-on +/- 
$30\degr$). For inclinations that are nearly face-on their separation quickly 
becomes unrealistically large for the interaction to take place and the 
projected shape becomes more circular. Our model is consistent with the latter 
in the sense that the projected surface becomes nearly circular in appearance 
when viewed face-on, and no longer reproduces the observed asymmetric shape. 
Even with the high-resolution of our WFPC2 images, we are unable to further 
constrain the range of inclinations for NML~Cyg's circumstellar envelope 
(compare the dot-dashed curve to the solid curve in Figure~\ref{surface}). 
Figure~\ref{nml_cygob2} shows the relative positions on the sky between 
NML~Cyg and Cyg~OB2. The relatively large angular size of Cyg~OB2 as viewed 
from NML~Cyg means that the incident UV flux is not truly plane parallel. The 
net effect of this would be to reshape the dissociation surface in 
Figure~\ref{surface}; making it {\it more} pointed along the direction of the 
incident flux, though the projected shape would still be nearly circular for 
low inclinations. With regards to the separation between NML~Cyg and Cyg~OB2 
($2.74\degr$ on the sky), an inclination of $90\degr$ (edge-on) corresponds to 
about 80~pc at Cyg~OB2's distance. The range of inclinations above allows for 
a maximum linear separation of up to $\sim100$~pc, and grows to nearly 500~pc 
for an inclination of $10\degr$. As an aside, if NML~Cyg is assumed to be 
closer than Cyg~OB2, then it's distance could be as little as $\sim1600$~pc. 
This is near the high end of \citet{danchi01}'s independent distance estimate 
of 1220 +/- 300~pc for NML Cyg based on Doppler-measured maser velocities.

It is worth mentioning that radiation and/or gas pressure are insufficient to 
mold the reflection nebula into the shape that we observe. However, we can 
show that the incident UV flux will destroy the dust grains embedded in the 
nebula, shaping their distribution in the nebula along the dissociation 
surface.

Stochastic heating is one method that is capable of heating the grains to 
sufficiently high temperatures for sublimation to occur. At a distance of 
$0\farcs1$ from NML~Cyg, the blackbody temperature of a grain will be about 
600~K, or even higher for non-perfect emitters. When a photon with sufficient 
energy is absorbed by a grain there will be a sharp increase in that grain's 
temperature that can be calculated from the heat capacity per volume for 
silicate grains: $C(T)/V = 3.41 \times 10^{7}$ ergs~K$^{-1}$cm$^{-3}$ 
\citep[for $T\geq500$~K]{draine85}. Dust destruction will take place if this 
increase is enough to raise the temperature above the sublimation limit. If we 
assume a rod-like geometry for the grains (more plausible than an idealized 
spherical grain), with a length $a$ and a length to thickness ratio of 
$\sim10$, then the rise in temperature from each photon is approximately:

\begin{eqnarray}
{\Delta}T \approx 1.2 \times 10^{-5} \left(\frac{E_{ph}}{eV}\right)\left(\frac{a}{\micron}\right)^{-3}
\end{eqnarray}

\noindent where $E_{ph}$ is the energy carried by each photon. A 70\AA~long 
grain that absorbs a 10~eV photon will increase its temperature by 400~K. 
Larger ($a\gg100$\AA) non-porous grains will not be significantly heated by 
this method. Even without such drastic heating, however, grain destruction may 
still take place slowly through other processes such as sputtering, charging 
(stripping the grain of charge with each photon absorption, leading to a net 
repulsive coulomb force that could break the grain apart), or chemical 
sputtering (erosion through chemical reaction with H, N, and O). These 
processes could erode the larger grains sufficiently that stochastic heating 
becomes significant. If the grain temperature in NML~Cyg's wind is 600~K or 
higher, then these combined processes will be efficient in breaking up the 
grains outside the photo-dissociation region.

The net result is a model for NML~Cyg's circumstellar nebula that has a 
layered structure of photo-dissociation boundaries. Outside each dissociation 
surface there are fewer molecules absorbing Cyg~OB2's near--UV photons, which 
leaves the dust grains unprotected, and thus more likely to be destroyed. The 
rapid decline of dust density outside the photo-dissociation region may 
explain the decline in the observed amount of  scattered light along the 
bean-shaped surface that is remarkably similar to the shape shown in 
Figure~\ref{surface}.

Alternatively, it is also possible that the near--UV photons are absorbed by 
the gas, heating it and resulting in a gas expansion front. The resulting 
acceleration (albeit small) would cause a drop in the density. If the dust 
were somehow accelerated by the gas (the reverse of what is normally expected 
for dust driven winds), then there would also be a corresponding drop in the 
dust density.

In Figure~\ref{nml2} we show the edge-on dissociation surface superimposed on 
our deconvolved F555W image. For display purposes, we placed the dissociation 
surface near the contour of the reflection nebula's edge with the axis of 
symmetry aligned with the direction of Cyg~OB2's center, $\sim288\degr$ E of N 
(at $\alpha$:~20$^{h}$33$^{m}$10$^{s}$, $\delta$:~+41\degr12$^{m}$, determined 
by \citealt{knodlseder00}). We assume that the embedded star is near the peak 
intensity in our images, which is located at 
$\alpha$:~20$^{h}$46$^{m}$25\fs573, $\delta$:~+40\degr07$^{m}$00\fs27 ({\it 
HST} WCS J2000). To get the best match when overlaying the dissociation 
surface on the optical image, we shifted the apparent position of the star by 
about 1 pixel to the North, indicated by the small white cross near the 
center. Figure~\ref{nml2} also shows the brightest of the integrated 22~GHz 
H$_{2}$O features superimposed on our F555W image. \citet{richards96} chose as 
their reference for NML~Cyg's location the integrated 22~GHz maser maximum. 
Therefore, we established the relative coordinate system offsets by aligning 
the strongest water vapor maser with the white cross. The spot size assigned 
to each maser is roughly proportional to its flux. The overall distribution of 
the masers matches the size of the bright nebulosity seen in our images, with 
the notable exception of the NW maser. There is some diffuse signal in the 
area around the NW maser in our deconvolved image, although we are unable to 
distinguish if the low-level signal is associated with the NW maser or is a 
residual PSF artifact that remains after deconvolution. Note that the NW/SE 
symmetry axis in the maser maps is not in the direction of Cyg~OB2.

It is especially interesting that the asymmetric one-sided distribution of the 
water masers is not only similar in extent to the reflection nebula, but also 
matches its convex shape. The dusty cocoon engulfing NML~Cyg must be the 
consequence of high mass loss in the RSG stage, but its envelope has most 
likely been shaped by its interaction with and proximity to Cyg~OB2. If the 
outflow from NML~Cyg is bipolar \citep{richards96}, then it appears that the 
molecular material SE of the star is preferentially shielded from 
photo-dissociation. Even without assuming bipolarity, there is more maser 
emission to the ESE, consistent with our model for NML~Cyg's circumstellar 
envelope.

\subsection{VX~Sgr}

The semi-regular supergiant VX~Sgr is a well studied OH/IR source whose 
spectral type varies from M4e~Ia to as late as M9.5~I \citep{humphreys72a}, 
and has been reported as late as M9.8 \citep{lockwood82}. The amplitude of the 
light variations in $V$ may reach 6~mag (\citeauthor{lockwood82}). Our images 
were taken when VX~Sgr was nearing maximum light in late 1999. Its distance is 
not well known, but assuming it is a member of the Sgr~OB1 association it is 
most likely in the inner spiral arm, between 1.5 to 2~kpc from the Sun 
\citep{humphreys72b}. Our images of VX~Sgr show that it is embedded in an 
envelope that is only slightly resolved (Figure~\ref{vxsgr}, compared to the 
WFPC2 PSF FWHM). There is little difference between our images, however, the 
F410M image was not saturated and so did not need to be PSF-patched. The 
extended envelope is nearly symmetric with a FWHM of $\sim0\farcs09$, 
approximately 150~AU at VX~Sgr's distance.

\citet{danchi94} estimate the inner radius of the dust shell to be at 
$\sim0\farcs06$. When combined with our measurements, we find that VX~Sgr's 
circumstellar envelope is a shell that is about 50~AU thick. Observations of 
the the H$_{2}$O and SiO masers show that they are $\sim0\farcs02$ from the 
star, whereas the OH masers extend out $\sim1\arcsec$ 
\citep{lane84,chapman86}.

\subsection{S~Per}

The RSG S~Per (M3-4e~Ia) is another well studied OH/IR source that 
is a known member of the Perseus OB1 association at a distance of 2.3~kpc 
\citep{humphreys78}. Our images show a star embedded in a circumstellar nebula 
(Figure~\ref{sper}). The envelope is $\sim0\farcs1$ across (FWHM), 
approximately 230~AU at S~Per's distance, and appears elongated along a NE/SW 
axis. Figure~\ref{sper} shows the F467M image, with the best PSF-patch, 
together with the WFPC2 PSF FWHM. The other images are very similar. The 
asymmetry in the envelope could be explained by bipolarity in the ejecta or a 
flattened circumstellar halo.

Recent OH and H$_{2}$O maser observations have shown an elongated structure 
around S~Per in good agreement with our images. Figure~1 from 
\citet{richards99} shows an elongated distribution for the integrated 22~GHz 
H$_{2}$O maser emission whose extent and symmetry axis match those seen in our 
images. The OH masers are more extended than the H$_{2}$O masers, and 
in \citeauthor{richards99}'s Figure~5 they show some indication of a NE/SW 
axis. However, the OH mainline masers are more clustered, as compared to the 
1612 MHz masers, with a handful of them more randomly dispersed, mainly to the 
North. \citet{vlemmings01} also find that the water masers have a similar 
distribution to that of \citeauthor{richards99}'s observations. Earlier work 
by \citet{diamond87} indicate a more E/W axis for water masers.

\section{The Intermediate--type Hypergiants}{\label{intermediate}}

Our imaging program also included three very luminous intermediate type 
hypergiant stars that are all well known for their records of spectroscopic 
and photometric variability.

$\rho$~Cas is the best known of this group and is famous for its historical 
and recent ``shell'' episodes 
\citep{bidelman57,beardsley61,boyarchuk88,lobel03} during which it temporarily 
develops TiO bands in a cool, optically thick wind with a very brief but high 
mass loss rate ($3\times10^{-2}~M_{\sun}$ in 200 days, \citealt{lobel03}). 
After each of these events the star quickly returns to its F supergiant 
spectrum. \citet{lobel03} showed that prior to its recent episode (2000-01), 
$\rho$~Cas displayed photometric variability indicative of  pulsational 
instability. The first infrared observations surprisingly did not show any 
evidence for dust, but IRAS observations showed that dust had formed sometime 
between 1973 and 1983 \citep{jura90}, in the expanding and cooling gas 
presumably from its 1946 episode. By 1989 the IR excess had weakened due to 
the dissipation and cooling of the grains as the shell expanded. With an 
expansion velocity of 35--40~km~s$^{-1}$ the shell should now be at 
$\sim0\farcs2$ from the star at its approximate distance of 2.5~kpc.

HR~8752`s atmospheric variations have been thoroughly discussed in a series of 
papers by de~Jager and collaborators 
\citep{dejager97a, dejager97b, dejager98}. It has a very high mass loss rate 
and apparent temperature changes of 1000~K or more \citep{israelian99}, 
possibly due to small changes in its wind or envelope. Like $\rho$~Cas it does 
not have a large IR excess, nor is it known to have any molecular emission, 
although circumstellar CO lines have been reported.

HR~5171a is spectroscopically similar to HR 8752, but has a very prominent 
10~$\mu$m silicate feature \citep{humphreys71}. Visual photometry of HR~5171a 
shows that it has been slowly getting fainter and redder \citep{vangenderen92} 
perhaps due to increased opacity in its wind or increased obscuration by dust.

Though these stars were excellent candidates to search for circumstellar 
material, we detect no evidence for shells or other ejecta. Our PSF 
subtractions and deconvolutions have a detection limit of approximately 5--7.5 
magnitudes, or about 100 to 1000 times, fainter than the stars at an angular 
separation of about $0\farcs1-0\farcs2$. Farther from the star the detection 
limit decreases, as there is less light in the wings of the PSF that can 
overwhelm any possible faint signal, and the background of the WFPC2 PC images 
are dominated by read noise $\sim1-3\arcsec$ from the stars. The angular 
radius from the star where the PSF contributes no more light, and read noise 
dominates, is given as $r_{RN}$ in Tables~\ref{limitsI} and \ref{limitsII}. 
The brightness limits ($I_{\lambda~max}$) corresponding to the read noise, 
along with the radius at which these limits take effect, are given in 
Table~\ref{limitsI} for these stars and Table~\ref{limitsII} for the M 
supergiants in Section~\ref{mtype}. Of course, it may be that any nebulosity 
around these stars is too faint to be detected with our methods.

We have also estimated how long it would take a hypothetical shell of material 
ejected with constant velocity (e.g. in a shell outburst, or other episodic 
high mass loss event) to reach a separation of $r_{RN}$ from the star 
(Table~\ref{time}). These expansion times are calculated for both a typical 
hypergiant wind velocity ($v_{exp}$) and a typical velocity for the violet 
wing of a P-Cygni profile ($v_{\infty}$) 
\citep[e.g.~see][]{lobel03,israelian99}. The expansion time estimates in 
Table~\ref{time} indicate that $\rho$~Cas, HR~8752, and HR~5171a have been 
losing mass at prodigious rates for no more than about $10^{3}$~yrs. We find 
no evidence for high mass loss events prior to 500-1000~yrs ago for these 
stars. If there has been any since, then the ejecta is too faint to detect 
given our detection limits.

\section{Mass Loss Histories and Evolutionary Status}{\label{massloss}}

Our original goal for this program of high resolution imaging and spectroscopy 
of the cool hypergiant stars using both the {\it HST} and ground based 
telescopes was to explore the evolutionary status and mass loss histories of 
these very luminous and unstable stars. Only the post-RSG IRC~+10420 
\citep{humphreys97,humphreys02} and the powerful OH/IR M supergiant VY~CMa 
\citep{smith01,smith04,humphreys05} have extensive circumstellar ejecta that 
has obviously been formed over hundreds of years including discrete high mass 
loss events.

As mentioned in the introduction, most of the stars on this program were 
selected because of their evidence for high mass loss and observed 
instabilities. Compared with VY~CMa and IRC~+10420, our results for the 
M--type supergiants were somewhat surprising. The relatively normal but highly 
luminous M supergiant $\mu$~Cep was on our observing program primarily for 
comparison with the high mass losing OH/IR M supergiants, S~Per, VX~Sgr and 
NML~Cyg. $\mu$~Cep's point-source appearance suggests that it has not yet 
entered the high mass loss and high dust formation period represented by the 
OH/IR sources that may occur near the end of their RSG stage. The 
circumstellar envelopes of VX~Sgr and S~Per are comparable to the expected 
extent of the region where dust would form for each star. NML~Cyg has more 
significant nebulosity, apparently shaped by the UV radiation of the nearby 
hot stars in Cyg~OB2. NML~Cyg is optically obscured, but it is in a unique 
environment. It is possible that NML~Cyg's circumstellar material has been 
largely dissipated by the winds and radiation pressure inside the Cygnus~X 
bubble after being photo-ionized and dissociated by the UV radiation from 
Cyg~OB2. If NML~Cyg were not located in such close proximity to Cyg~OB2, it 
might show a much more extended nebula comparable to VY~CMa.

NML~Cyg, VX~Sgr, S~Per, and VY~CMa are all very luminous RSGs  and 
strong OH/IR sources with high mass loss rates. However, VY~CMa's mass loss 
rate is about ten times higher than that of the other three stars. Thus, it 
may be in a unique high mass loss stage that the others will eventually pass 
through. Given the resemblance of VY~CMa's circumstellar environment to that 
of IRC~+10420, it is tempting to think that VY~CMa is emerging from the 
optically thick cocoon that hides many of the OH/IR stars and is about to 
leave the RSG region on a blueward track as a yellow hypergiant, although 
there are questions about the origin of VY~CMa's diffuse nebulosity, see 
\citealt{humphreys05}. But then, why do the other yellow hypergiants, such as 
$\rho$~Cas, not have circumstellar nebulae if they have formerly been red 
supergiants with high mass loss episodes as OH/IR supergiants?

At this time, our results suggest that $\rho$~Cas, HR~8752, and HR~5171a 
appear to be point-sources and there is no evidence for distant nebulosity. 
However, these stars are in a dynamically unstable region of the HR diagram 
\citep{dejager98,humphreys02}. Their high mass loss rates 
($\sim10^{-5}~M_{\sun}yr^{-1}$) and spectroscopic and photometric variability 
confirm that they are unstable \citep{israelian99}, although their continuous 
mass loss rates are also about 10 times less than in IRC~+10420. The lack of 
readily apparent ejecta associated with these stars suggests that if they are 
post-RSGs, they have only recently encountered this unstable region in their 
blueward evolution, and a star like $\rho$~Cas with its multiple shell 
ejection episodes may eventually resemble IRC~+10420. However, the lack of 
more distant nebulosity or fossil shells from a previous RSG state like that 
of S~Per or VX~Sgr suggests that there has either been sufficient time for the 
material to dissipate or perhaps an extensive nebula like that around VY~CMa, 
or even NML~Cyg, never formed. We also note that these three stars are 
somewhat less luminous than IRC~+10420 (and VY~CMa), and therefore may have a 
lower initial mass leading to a net lower mass loss rate. We also must 
consider the possibility that the extensive nebulae and the evidence for 
discrete and localized high mass loss events observed in the ejecta of VY~CMa 
and IRC~+10420 may be triggered by instabilities encountered at their somewhat 
higher masses and luminosities; that is, in stars closer to the upper 
luminosity boundary.

\acknowledgments

We are pleased to acknowledge interesting conversations with Kris Davidson and 
Michael Jura on the physics of circumstellar nebulae, and with David Boboltz 
regarding his recent VLA observations. We also thank Nathan Smith for advice 
on our analysis and interpretation of the WFPC2 images. This work is based on 
observations made with the NASA/ESA {\it Hubble Space Telescope}, obtained at 
the Space Telescope Science Institute, which is operated by the Association of 
Universities for Research in Astronomy, Inc., under NASA contract NAS5-26555.

\clearpage

\begin{figure}
\epsscale{1.0}
\plotone{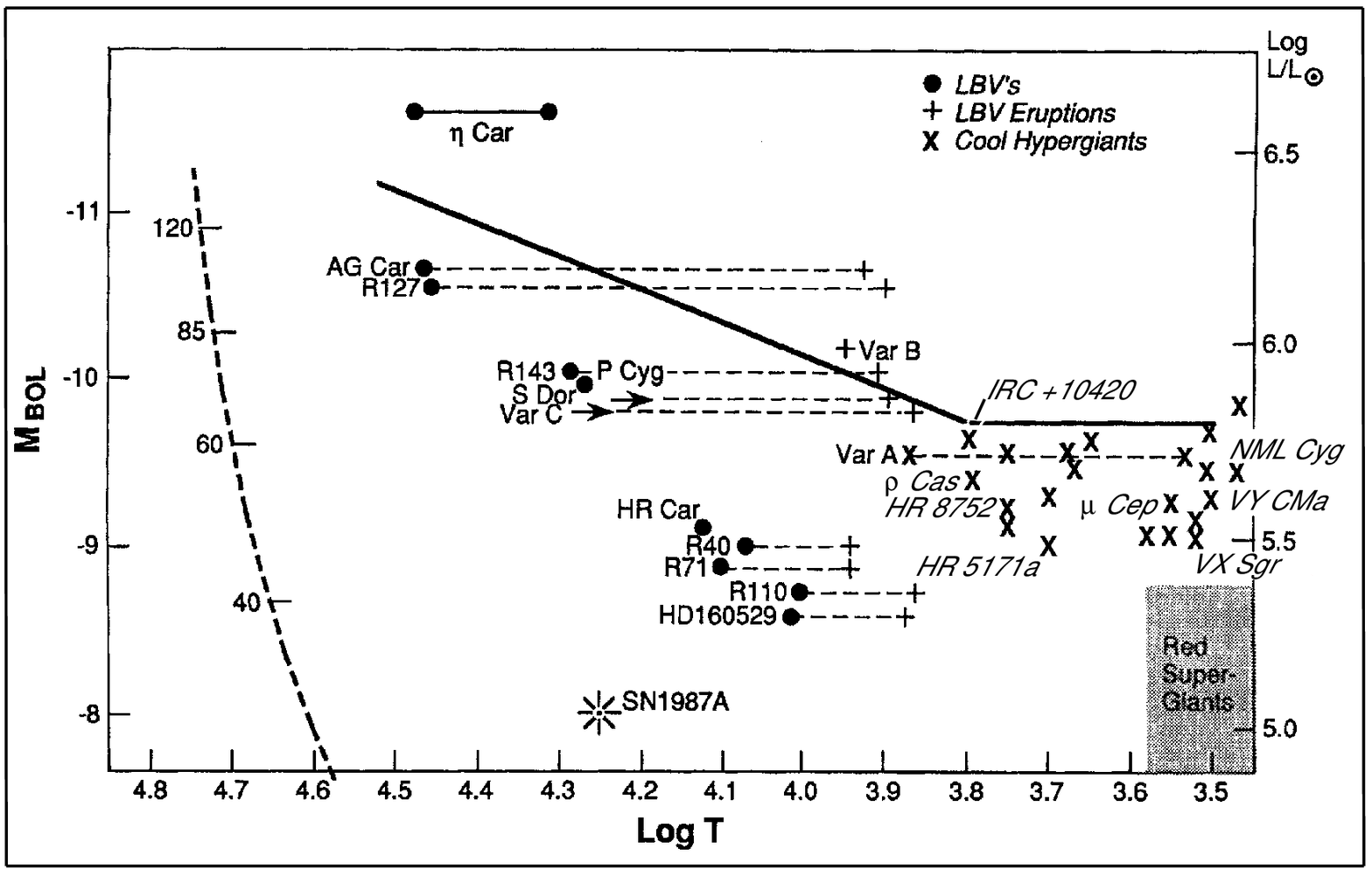}
\caption{A schematic HR Diagram for the most luminous stars in Local Group 
galaxies. The empirical upper luminosity boundary is shown as a solid line, 
and the cool hypergiants are labeled with X's.
\label{hrd}}
\end{figure}

\clearpage

\begin{figure}
\epsscale{0.67}
\plotone{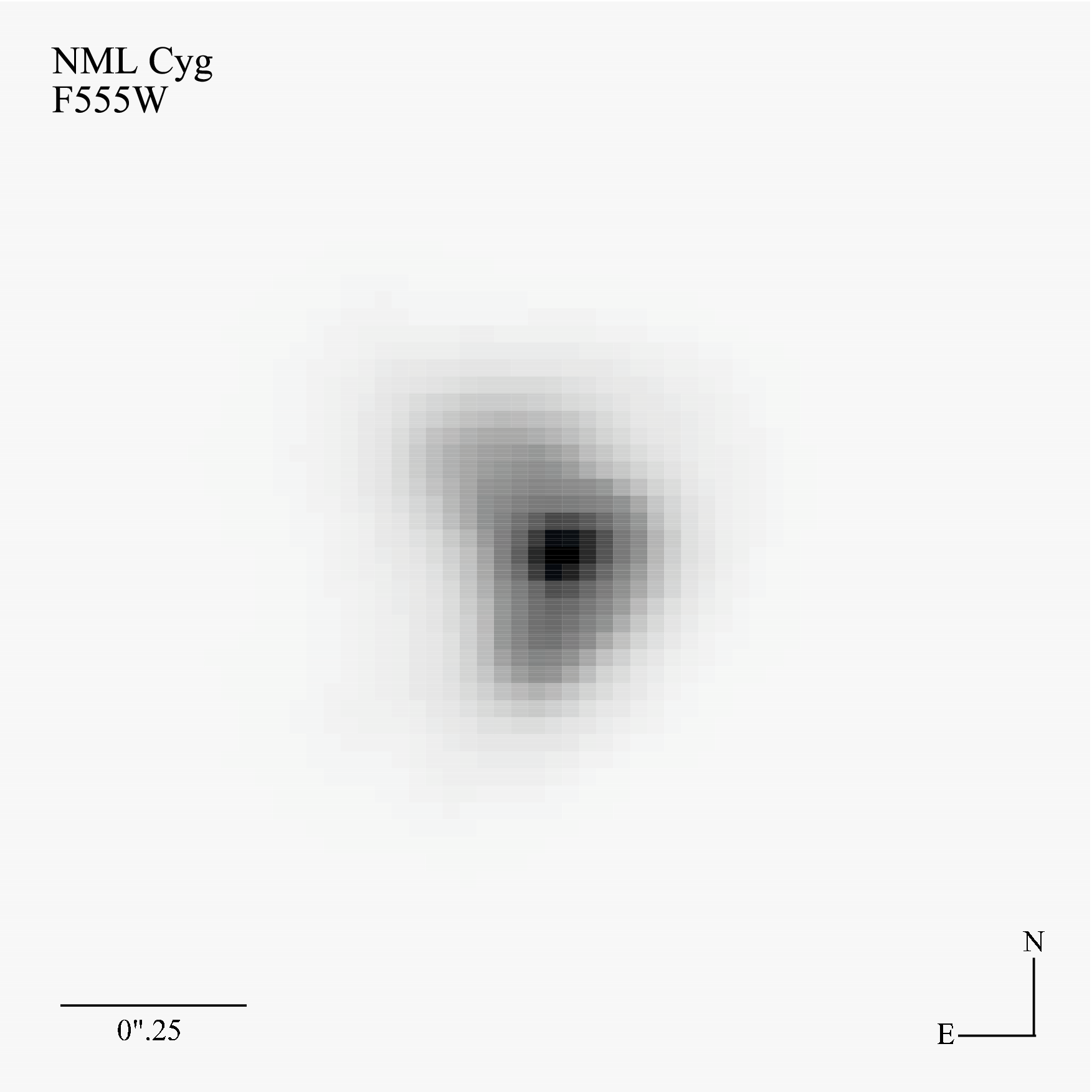}
\caption{Our deepest image of NML~Cyg shows its envelope has a peculiar 
asymmetric shape.
\label{nml1}}
\end{figure}

\clearpage

\begin{figure}
\epsscale{0.67}
\plotone{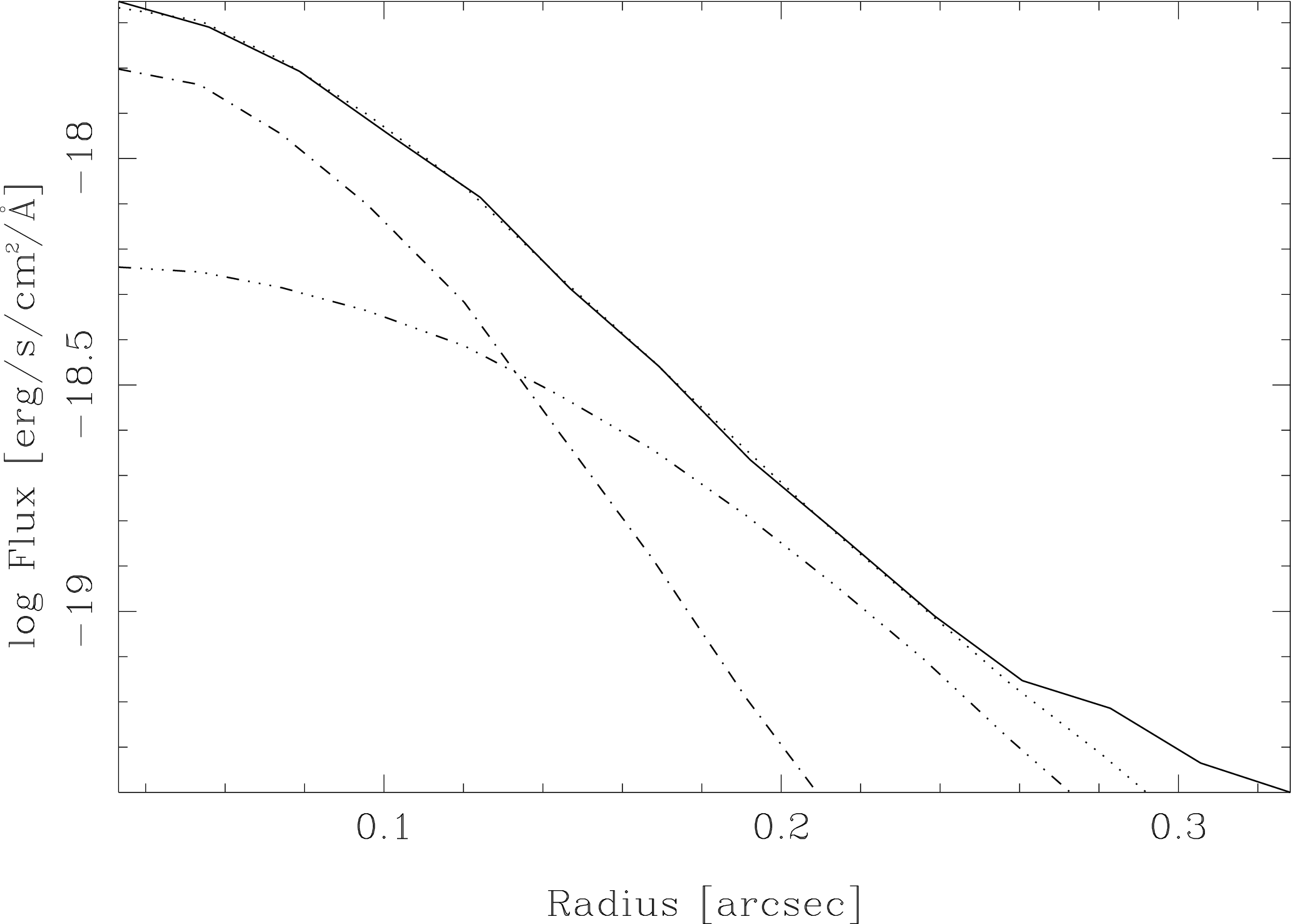}
\caption{The solid curve is a radial profile of the image in 
Figure~\ref{nml1}, in a $60\degr$ wedge in the direction of Cyg~OB2. We find 
that two components are necessary to fit the observed shape of the profile, 
one for the embedded star (dot-dashed curve - a gaussian, FWHM=4.7~pix, 
convolved with the PSF) and another for the asymmetric nebula (dot-dot-dashed 
curve - a gaussian, FWHM=9.4~pix, convolved with the PSF). The latter can be 
used as an estimate for the size scale for the dissociation surfaces, $r_{o}$. 
The dotted curve is the two components combined. The bump in the solid curve 
at $\sim0\farcs3$ is the diffuse light that may be coincident with the NW 
maser observed by \citet{richards96}.
\label{nmlfit}}
\end{figure}

\clearpage

\begin{figure}
\epsscale{0.67}
\plotone{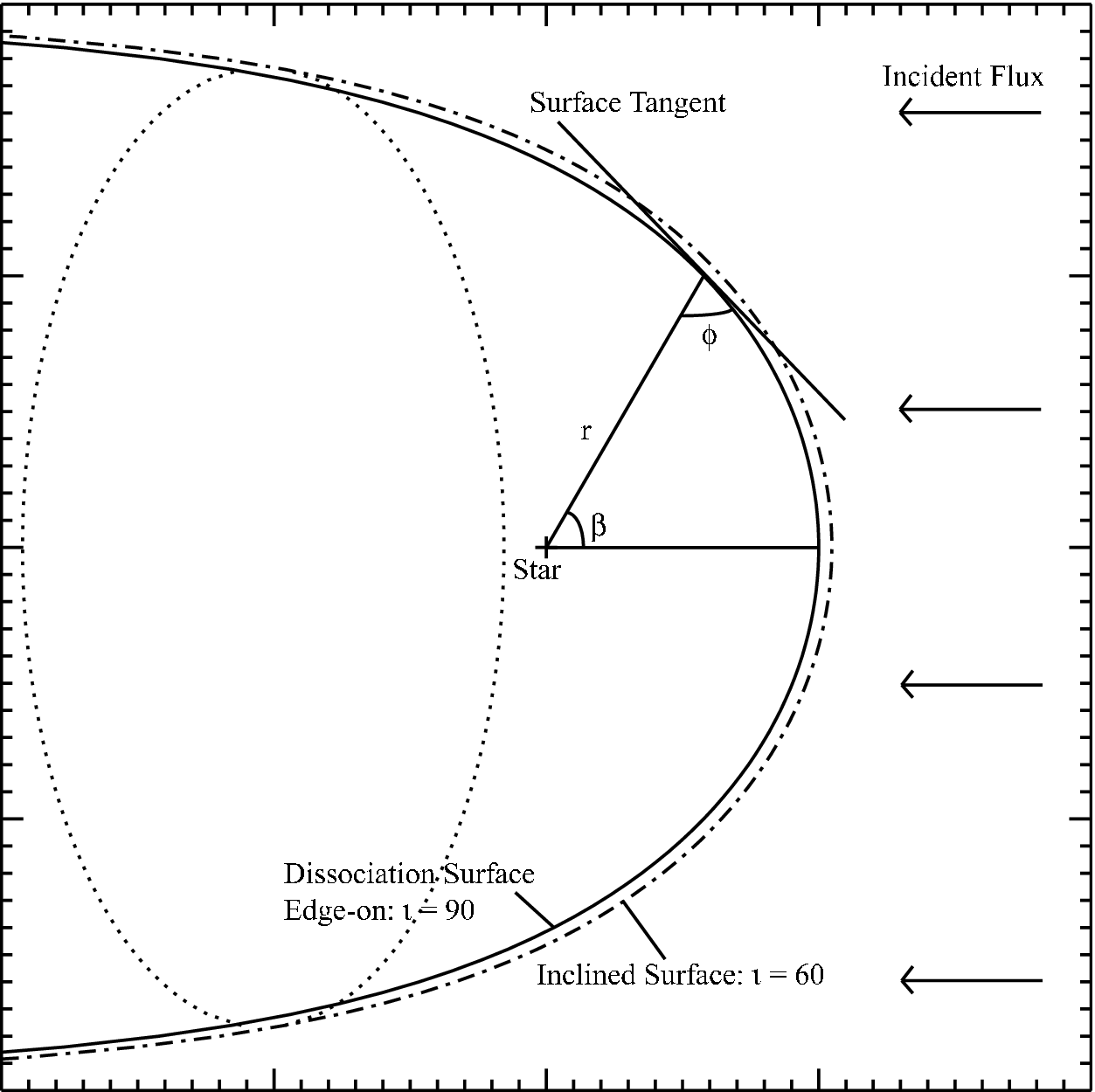}
\caption{Schematic of the interaction between the spherically symmetric 
expanding envelope from NML Cyg and the UV flux from Cyg OB2. The shape of the 
dissociation surface is calculated from eq. (3). Note that the location of the 
star relative to this surface is fixed and is invariant with changes in the 
size of the dissociation surface, as set by the distance $r_{o}$. The solid 
curve represents the surface as seen edge-on (inclination $90\degr$), and the 
dot-dashed curve (with dotted circle) represents the same surface seen with an 
inclination of $60\degr$ to our line-of-sight.
\label{surface}}
\end{figure}

\clearpage

\begin{figure}
\epsscale{0.67}
\plotone{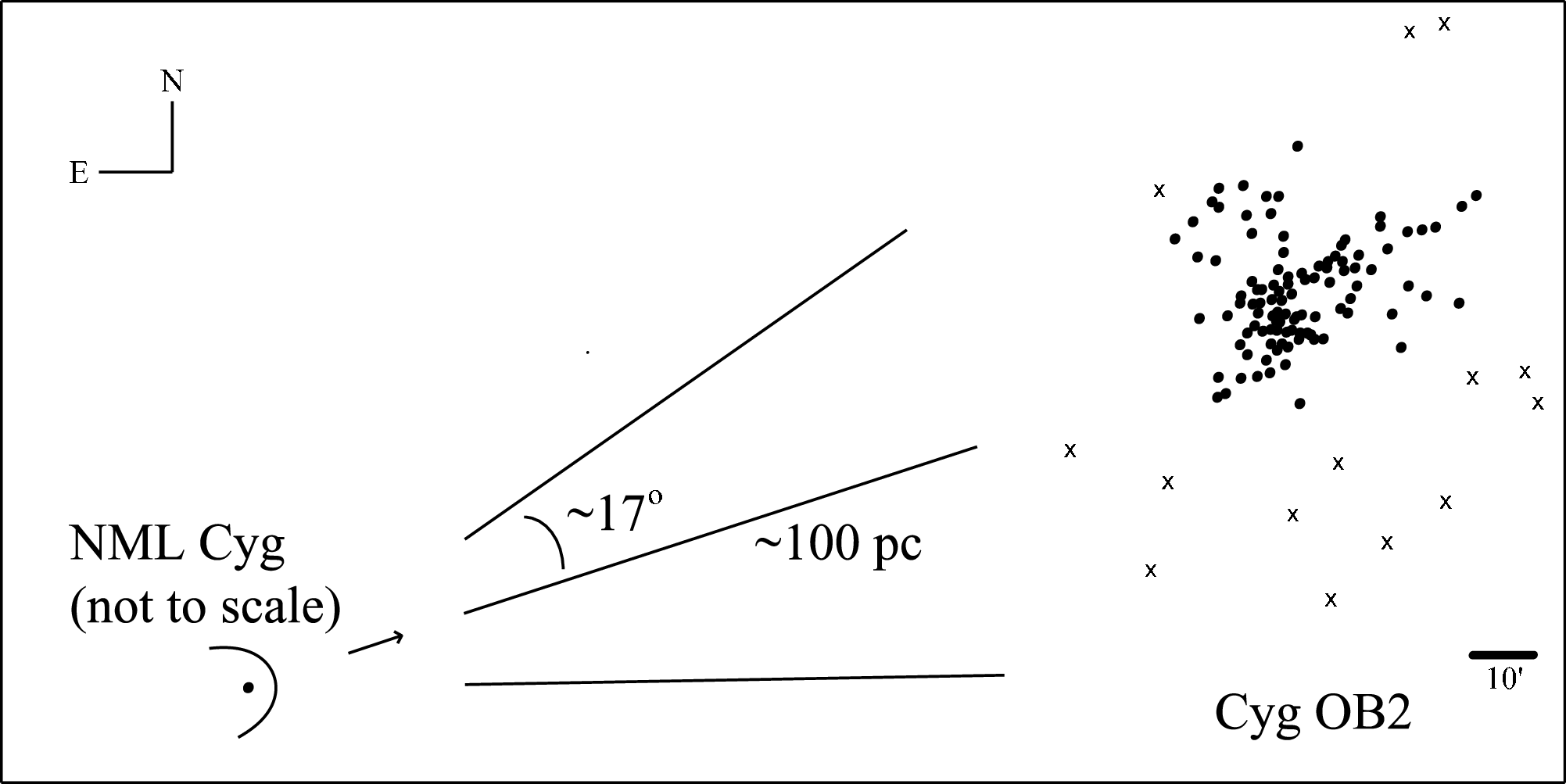}
\caption{This figure indicates the proximity of NML~Cyg to the Cyg~OB2 
association. NML~Cyg is at least 80~pc from Cyg~OB2, with a distance of 
1.7~kpc. If NML~Cyg is inclined w.r.t. our line-of-sight, then the distance 
between them is greater. The schematic of Cyg~OB2 is adapted from 
\citet{hanson03}'s Figure~4; known members are shown as circles, and recently 
confirmed members are shown as x's. The relatively close proximity implies 
that the wind from Cyg~OB2 is probably not plane parallel. NML~Cyg, and the 
dissociation surface (from Figure~\ref{surface}), are not to scale.
\label{nml_cygob2}}
\end{figure}

\clearpage

\begin{figure}
\epsscale{0.67}
\plotone{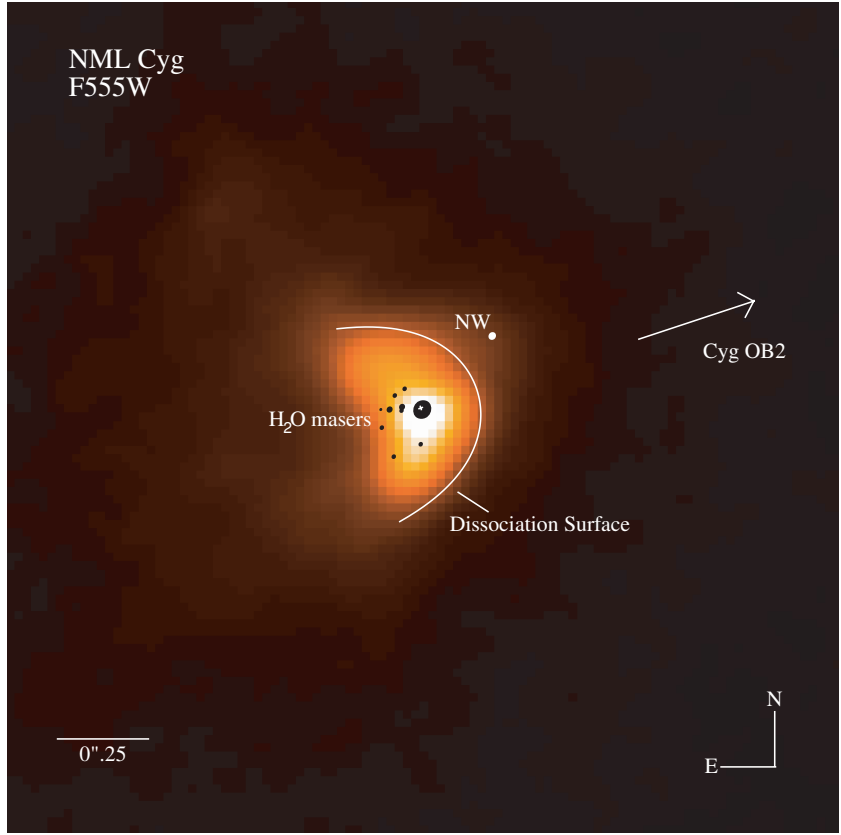}
\caption{Figure~\ref{nml1} after deconvolution shows that NML~Cyg's obscuring 
envelope is oriented with its symmetry axis in the direction of the nearby 
Cyg~OB2 association. We have superimposed the brightest 22~GHz H$_{2}$O maser 
features from Figure 1 of \citet{richards96}, and the edge-on dissociation 
surface from Figure~\ref{surface}. The size of the masers is roughly 
proportional to their flux. The star is shown by the white mark and large 
maser near the center. We have centered the superimposed features about one 
pixel to the North of the peak in the image. For display the dissociation 
surface is placed along the outer edge of the envelope as seen in this image. 
The envelope is likely shaped by photo-dissociation of the surrounding 
molecular material by the UV wind from the nearby Cyg~OB2 association. This 
image is displayed with a square root scale.
\label{nml2}}
\end{figure}

\clearpage

\begin{figure}
\epsscale{0.67}
\plotone{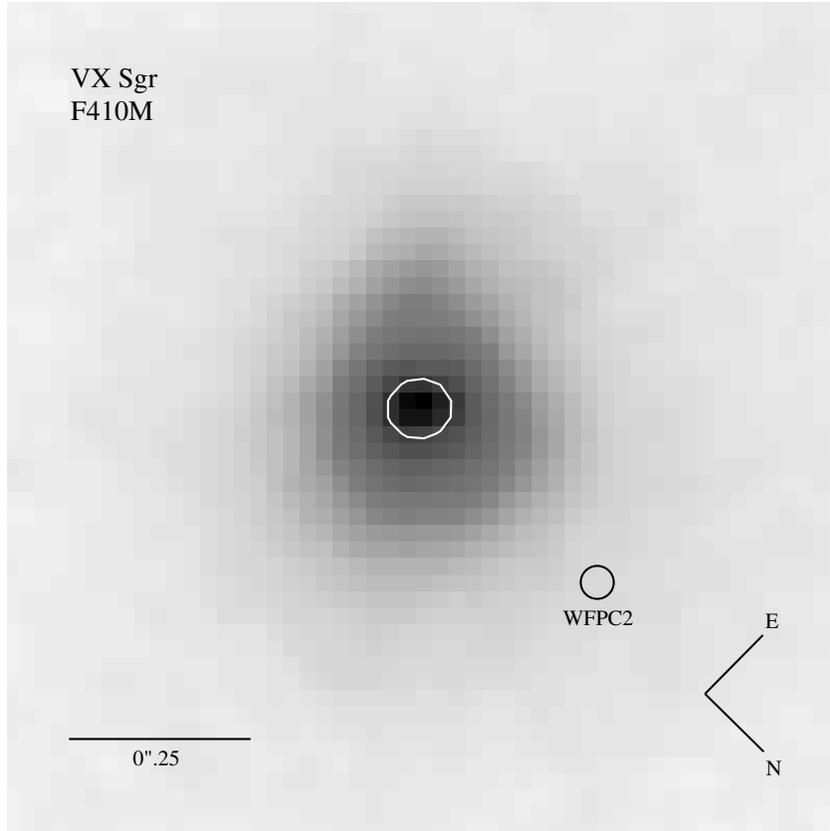}
\caption{This image shows VX~Sgr's extended envelope compared with the WFPC2 
PSF FWHM (black circle). The envelope is $\sim0\farcs09$ across (FWHM, white 
contour), approximately 150~AU at 1.7~kpc. The slight bulge at the top of the 
star is probably the result of a small amount of bleeding in the CCD. This 
image is displayed with a square root scale.
\label{vxsgr}}
\end{figure}

\clearpage

\begin{figure}
\epsscale{0.67}
\plotone{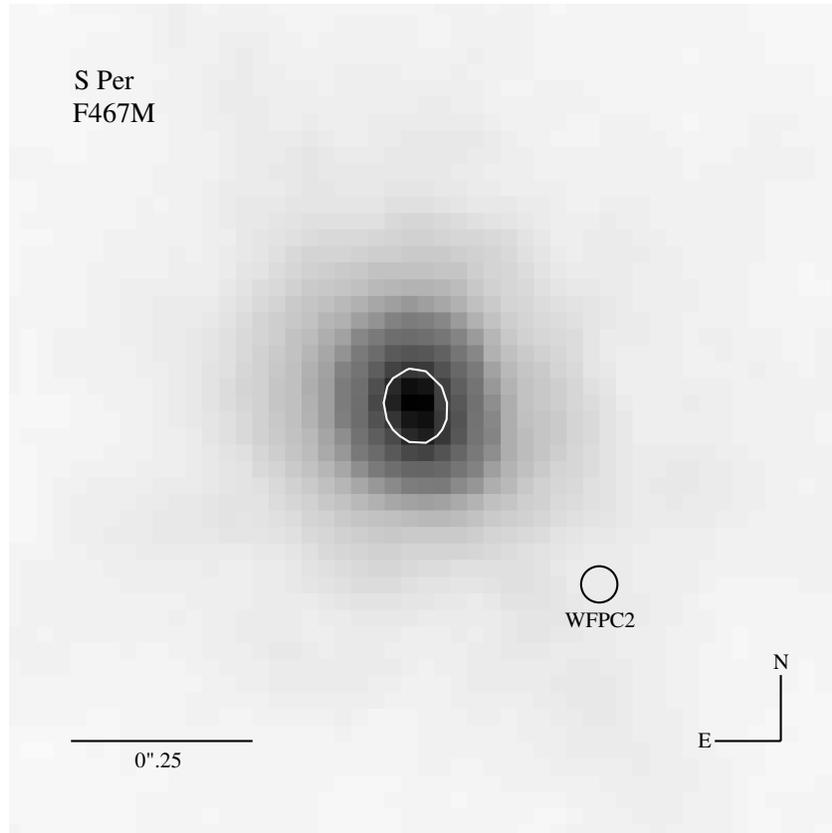}
\caption{This image shows S~Per's extended envelope compared with the WFPC2 
PSF FWHM (black circle). The envelope is $\sim0\farcs1$ across (FWHM, white 
contour), approximately 230~AU at 2.3~kpc, with a NE/SW alignment. The 
elongated shape may be due to bipolar ejecta or a flattened circumstellar 
halo. This image is displayed with a square root scale.
\label{sper}}
\end{figure}

\clearpage

\renewcommand{\arraystretch}{0.7}
\begin{deluxetable}{lcrrl}
\tabletypesize{\scriptsize}
\tablecaption{HST/WFPC2 PC Observations \label{data}}
\tablewidth{0pt}
\tablehead{
Star & \colhead{Filter} & \colhead{$\bar{\lambda}$(\AA)\tablenotemark{a}} & 
\colhead{$\delta\bar{\lambda}$(\AA)\tablenotemark{a}} & \colhead{Exposures~(s)}
}
\startdata
NML~Cyg & F439W & 4292.6 & 473.2 & 6$\times$500\tablenotemark{d} \\
 & F555W & 5336.8 & 1228.4 & 20, 100, 4$\times$400 \\
 & F656N & 6563.8 & 21.5 & 20, 2$\times$260 \\
 & F675W & 6677.4 & 866.8 & 0.5, 10 \\
 & & & & \\
VX~Sgr & F410M & 4085.7 & 146.8 & 2$\times$60 \\
 & F467M & 4667.7 & 166.5 & 10, 2$\times$60\tablenotemark{c} \\
 & F547M & 5467.8 & 483.2 & 0.23, 2$\times$6\tablenotemark{c} \\
 & F656N & 6563.8 & 21.5 & 5\tablenotemark{c} \\
 & & & & \\
S~Per & F410M & 4085.7 & 146.8 & 20, 2$\times$200\tablenotemark{c} \\
 & F467M & 4667.7 & 166.5 & 3, 2$\times$30\tablenotemark{c} \\
 & F547M & 5467.8 & 483.2 & 0.11, 2$\times$5\tablenotemark{c} \\
 & & & & \\
$\mu$~Cep & F375N & 3732.2 & 24.4 & 10, 2$\times$100\tablenotemark{c} \\
 & F437N & 4369.1 & 25.2 & 0.5, 2$\times$5\tablenotemark{c} \\
 & F502N & 5012.4 & 26.9 & 0.11, 1\tablenotemark{c}, 2$\times$5\tablenotemark{c} \\
 & F656N & 6563.8 & 21.5 & 0.11\tablenotemark{c} \\
 & & & & \\
HR~5171a & F375N & 3732.2 & 24.4 & 2$\times$300\tablenotemark{b} \\
 & F437N & 4369.1 & 25.2 & 2, 2$\times$26 \\
 & F502N & 5012.4 & 26.9 & 0.2, 2$\times$5\tablenotemark{c} \\
 & F656N & 6563.8 & 21.5 & 0.5\tablenotemark{c} \\
 & & & & \\
HR~8752 & F375N & 3732.2 & 24.4 & 5, 2$\times$80\tablenotemark{c} \\
 & F437N & 4369.1 & 25.2 & 0.11, 2$\times$5\tablenotemark{c} \\
 & F502N & 5012.4 & 26.9 & 0.11, 2$\times$1.4\tablenotemark{c} \\
 & F656N & 6563.8 & 21.5 & 0.11, 0.5\tablenotemark{c} \\
 & & & & \\
$\rho$~Cas & F375N & 3732.2 & 24.4 & 5, 2$\times$30\tablenotemark{c} \\
 & F437N & 4369.1 & 25.2 & 0.2, 2$\times$10\tablenotemark{c} \\
 & F502N & 5012.4 & 26.9 & 0.11, 1\tablenotemark{c}, 2$\times$5\tablenotemark{c} \\
 & F656N & 6563.8 & 21.5 & 0.11, 1\tablenotemark{c} \\
\enddata

\tablenotetext{a}{\citet{biretta00}}
\tablenotetext{b}{No dithering}
\tablenotetext{c}{Saturated}
\tablenotetext{d}{No detection}

\end{deluxetable}
\renewcommand{\arraystretch}{1}

\clearpage

\renewcommand{\arraystretch}{1.0}
\begin{deluxetable}{lccccccccccc}
\tabletypesize{\scriptsize}
\tablecaption{Circumstellar Detection Limits I\label{limitsI}}
\tablewidth{0pt}
\tablehead{
 & \multicolumn{2}{c}{$\rho$~Cas} && \multicolumn{2}{c}{HR~8752} && \multicolumn{2}{c}{HR~5171a} && \multicolumn{2}{c}{$\mu$~Cep}\\
 & \colhead{$r_{RN}$} & \colhead{$I_{\lambda~max}$\tablenotemark{a}} && \colhead{$r_{RN}$} & \colhead{$I_{\lambda~max}$\tablenotemark{a}} && \colhead{$r_{RN}$} & \colhead{$I_{\lambda~max}$\tablenotemark{a}} && \colhead{$r_{RN}$} & \colhead{$I_{\lambda~max}$\tablenotemark{a}}
}
\startdata
F656N & 3\farcs2 & 1.2$\times10^{-2}$ && 1\farcs6 & 1.9$\times10^{-2}$ && 1\farcs4 & 1.9$\times10^{-2}$ && 2\farcs3 & 8.5$\times10^{-2}$ \\
F502N & 3\farcs4 & 2.7$\times10^{-3}$ && 3\farcs0 & 9.3$\times10^{-3}$ && 1\farcs4 & 2.6$\times10^{-3}$ && 3\farcs2 & 2.7$\times10^{-3}$ \\
F437N & 2\farcs3 & 3.3$\times10^{-3}$ && 2\farcs3 & 6.4$\times10^{-3}$ && 1\farcs1 & 1.2$\times10^{-3}$ && 2\farcs3 & 6.7$\times10^{-3}$ \\
F375N & 1\farcs7 & 4.2$\times10^{-3}$ && 3\farcs0 & 1.6$\times10^{-3}$ && 0\farcs9 & 3.1$\times10^{-4}$ && 1\farcs6 & 1.2$\times10^{-3}$ \\
\enddata

\tablenotetext{a}{ergs~sec$^{-1}$cm$^{-2}${\AA}$^{-1}$sr$^{-1}$}

\end{deluxetable}
\renewcommand{\arraystretch}{1}

\renewcommand{\arraystretch}{1.0}
\begin{deluxetable}{lcccccccc}
\tabletypesize{\scriptsize}
\tablecaption{Circumstellar Detection Limits II\label{limitsII}}
\tablewidth{0pt}
\tablehead{
 & \multicolumn{2}{c}{NML~Cyg} && \multicolumn{2}{c}{VX~Sgr} && \multicolumn{2}{c}{S~Per}\\
 & \colhead{$r_{RN}$} & \colhead{$I_{\lambda~max}$\tablenotemark{a}} && \colhead{$r_{RN}$} & \colhead{$I_{\lambda~max}$\tablenotemark{a}} && \colhead{$r_{RN}$} & \colhead{$I_{\lambda~max}$\tablenotemark{a}}
}
\startdata
F656N & 0\farcs8 & 1.9$\times10^{-5}$ && 1\farcs6 & 1.9$\times10^{-3}$ && \nodata & \nodata \\
F675W & 0\farcs9 & 1.8$\times10^{-5}$ && \nodata & \nodata && \nodata & \nodata \\
F555W & 2\farcs7 & 2.0$\times10^{-7}$ && \nodata & \nodata && \nodata & \nodata \\
F547M & \nodata & \nodata && 2\farcs8 & 5.7$\times10^{-5}$ && 2\farcs6 & 6.9$\times10^{-5}$ \\
F467M & \nodata & \nodata && 2\farcs4 & 3.1$\times10^{-5}$ && 1\farcs9 & 6.3$\times10^{-5}$ \\
F410M & \nodata & \nodata && 1\farcs3 & 5.6$\times10^{-5}$ && 1\farcs9 & 3.4$\times10^{-5}$ \\
\enddata

\tablenotetext{a}{ergs~sec$^{-1}$cm$^{-2}${\AA}$^{-1}$sr$^{-1}$}

\end{deluxetable}
\renewcommand{\arraystretch}{1}

\clearpage

\renewcommand{\arraystretch}{1.0}
\begin{deluxetable}{lccccc}
\tabletypesize{\scriptsize}
\tablecaption{Parameters for the Intermediate-Type Hypergiant Detection Limits\label{time}}
\tablewidth{0pt}
\tablehead{
 & $\rho$~Cas && HR~8752 && HR~5171a
}
\startdata
Distance~(kpc)\tablenotemark{a} & 2.5 && 3.5 && 3.3 \\
$r_{RN}$~(\arcsec) & 1.7--3.4 && 1.6--3.0 && 0.9--1.4 \\
$v_{exp}$~(km~s$^{-1}$) & 35 && 35 && 35 \\
$t_{exp}$~(yr) & 575--1150 && 760--1420 && 400--625 \\
$v_{\infty}$~(km~s$^{-1}$) & 100 && 100 && 100 \\
$t_{\infty}$~(yr) & 200--400 && 265--500 && 140--220 \\
\enddata

\tablenotetext{a}{\citet{humphreys78}}
\tablecomments{The expansion times are for a hypothetical shell of material ejected with constant velocity to reach a separation of $r_{RN}$ from the star at which we measure the detection limits given in Tables \ref{limitsI} and \ref{limitsII}. The times $t_{exp}$ and $t_{\infty}$ correspond to typical hypergiant wind velocities and P-Cygni profile velocities, respectively. }

\end{deluxetable}
\renewcommand{\arraystretch}{1}

\end{document}